\documentclass[prb,twocolumn,showpacs,byrevtex,floatfix,superscriptaddress]{revtex4}
\usepackage{amssymb,graphicx,afterpage,ulem}
\begin{document}
\title{\boldmath Experimental band structure of the nearly half-metallic CuCr$_2$Se$_4$: An optical and magneto-optical study\unboldmath}
%
% authors and affiliations
%
%
\author{S. Bord\'acs}
\affiliation{Department of Physics, Budapest University of
Technology and Economics and Condensed Matter Research Group of the Hungarian Academy of Sciences, 1111 Budapest, Hungary} \affiliation{Multiferroics Project, ERATO, Japan Science and Technology Agency (JST), Japan c/o Department of Applied Physics, The University of Tokyo, Tokyo 113-8656, Japan}
\author{I. K\'ezsm\'arki}
\affiliation{Department of Physics, Budapest University of
Technology and Economics and Condensed Matter Research Group of the Hungarian Academy of Sciences, 1111 Budapest, Hungary} \affiliation{Multiferroics Project, ERATO, Japan Science and Technology Agency (JST), Japan c/o Department of Applied Physics, The University of Tokyo, Tokyo 113-8656, Japan} \affiliation{Department of Applied
Physics, University of Tokyo, Tokyo 113-8656, Japan}
\author{K. Ohgushi}
\affiliation{Institute for Solid State Physics, University of Tokyo, Kashiwanoha, Kashiwa, Chiba 277-8581, Japan}
\author{Y. Tokura}
\affiliation{Multiferroics Project, ERATO, Japan Science and Technology Agency (JST), Japan c/o Department of Applied Physics, The University of Tokyo, Tokyo 113-8656, Japan}
\affiliation{Department of Applied Physics, University of Tokyo,
Tokyo 113-8656, Japan}
\date{\today}
\begin{abstract}
Diagonal and off-diagonal optical conductivity spectra have been
determined form the measured reflectivity and magneto-optical Kerr
effect (MOKE) over a broad range of photon energy in the itinerant
ferromagnetic phase of CuCr$_2$Se$_4$ at various temperatures down
to T=10\,K. Besides the low-energy metallic contribution and the
lower-lying charge transfer transition at $E$$\approx$$2$\,eV, a
sharp and distinct optical transition was observed in the
mid-infrared region around $E$$=$$0.5$\,eV with huge
magneto-optical activity. This excitation is attributed to a
parity allowed transition through the Se-Cr hybridization-induced
gap in the majority spin channel. The large off-diagonal
conductivity is explained by the high spin polarization in the
vicinity of the Fermi level and the strong spin-orbit interaction
for the related charge carriers. The results are discussed in
connection with band structure calculations.
\end{abstract}
%
% PACS numbers
%
\pacs{\ }
\maketitle

\section{INTRODUCTION}

The ferrimagnetic metal CuCr$_2$Se$_4$ shows the highest critical
temperature T$_c$=430\,K among chromium spinel chalcogenides.
\cite{Landolt} Although the lattice constants of the ferromagnetic
semiconductors CdCr$_2$S$_4$ and CdCr$_2$Se$_4$ differ only by
1-4$\%$, their ferromagnetism is considerably weakened by the
complete filling of the valance band as reflected by T$_c$=85\,K
and 129\,K, respectively \cite{Landolt}. CuCr$_2$Se$_4$ exhibits
large magneto-optical Kerr effect (MOKE) in the near infrared
photon energy region at room temperature which makes this compound
a promising candidate for magneto-optical devices. \cite{Reim91}
Materials from the same family show interesting magneto-transport
phenomena like the colossal magnetoresistance\cite{Ramirez97} in
Fe$_{1-x}$Cu$_{x}$Cr$_2$S$_4$, and the colossal
magnetocapacitance\cite{Loidl05} in CdCr$_2$S$_4$. Moreover,
recent band structure calculations indicate that CuCr$_2$Se$_4$ is
almost half-metallic,\cite{Antonov99,Sarma07,Butler08} and the
density of states for spin down electrons can be fully suppressed
with cadmium doping, i.e. a perfect half-metallic situation can be
realized. \cite{Butler08}

The strong ferrimagnetism in CuCr$_2$Se$_4$ was first explained by
Lotgering and Stapele assuming the mixed-valence state of
Cr$^{3+}$ and Cr$^{4+}$ with the monovalent Cu$^+$; thus, this
compound was classified as a d-metal with closed Se 4p shell.
\cite{Lotgering67} In this picture only the chromium sites are
magnetic and the double exchange mechanism between the Cr$^{3+}$
and the Cr$^{4+}$ ions align their magnetic moment parallel.
However, early neutron diffration studies indicated that each
chromium is in the Cr$^{3+}$ state. \cite{Colominas66} Later,
Goodenough proposed the copper ions to be divalent Cu$^{2+}$ and
as a source of the magnetism the 90 degree superexchange to be
responsible for the coupling between the Cr$^{3+}$ ions through
the completely filled Se 4p states. \cite{Goodenough69} The recent
XMCD measurements by Kimura et al.\cite{Kimura01} settled the long
standing issue of the valance state of CuCr$_2$Se$_4$. They have
confirmed the Cr$^{3+}$ state, however, they have found almost
monovalent copper and a delocalized hole in the Se 4p band with a
magnetic moment anti-parallel to the moment of the Cr$^{3+}$ ions.
Based on these experimental results Sarma et al.\cite{Sarma07}
have interpreted the ferrimagnetism in terms of a kinetic-energy
driven mechanism in which the hybridization between the localized
Cr$^{3+}$ ions and the delocalized Se 4p band results a hole
mediated exchange. Their density functional calculation indicates
the appearance of a hybridization induced hump-like structure at
the Fermi energy only for the up spin states in accordance with
other band structure calculations. \cite{Antonov99,Butler08}

In order to have a deeper understanding of the electronic
structure of CuCr$_2$Se$_4$ we have investigated the low-energy
(E=0.1-4\,eV) charge excitations over the temperature range of
T=10-300\,K by determining both diagonal and off-diagonal element
of the optical conductivity tensor. Beside the low-energy response
of the metallic carriers and the charge transfer excitations above
$E\gtrsim2$\,eV, we have found a sharp and distinct
optical transition in the mid-infrared region with large
magneto-optical activity.

\section{EXPERIMENTAL}

Single crystals of CuCr$_2$Se$_4$ were grown by the chemical vapor
transport method. Details of the preparation and the structure
characterization were given elsewhere. \cite{Ohgushi08} All the
optical measurements were carried out with nearly normal incidence
on the as-grown (111) surface. In order to determine the diagonal
optical conductivity, reflectivity spectra was measured over a
broad energy range (E=0.08-26\,eV and E=0.08-5\,eV at room and low
temperature, respectively) to facilitate the proper Kramers-Kronig
transformation. We have measured the complex magneto-optical Kerr
angle $\Phi_{Kerr}=\theta_{Kerr} +i\eta_{Kerr}$, which allows the
direct determination of the off-diagonal conductivity, between
E=0.12-4\,eV with a polarization modulation technique.
\cite{Sato81} In the mid-infrared region a Fourier transform
infrared spectrometer was combined with a ZnSe photoelastic
modulator (Hinds, II/ZS50) \cite{Ohgushi08} to perform
measurements down to as low energy as E=0.12\,eV, while above
$E>0.7$\,eV a CaF$_2$ photoelastic modulator (Hinds, I/CF50) and a
grating spectrometer was used. The external magnetic field
B=$\pm$0.25\,T was applied by a permanent magnet along the [111]
easy axis of the magnetization and also parallel to the
propagation direction of the light. In the above arrangement, the
conductivity tensor has the following form:
\begin{equation}
\uuline{\sigma} = \left[
\begin{array}{ccc}
\sigma_{xx} & \sigma_{xy} & 0\\
-\sigma_{xy} & \sigma_{xx} & 0 \\
0 & 0 & \sigma_{zz}
\end{array}
\right],
\end{equation}
in our notation the x, y, z directions do not correspond to the
main cubic axes, as z is chosen parallel to the [111] easy axis.

\begin{figure}[t!]
\includegraphics[width=3in]{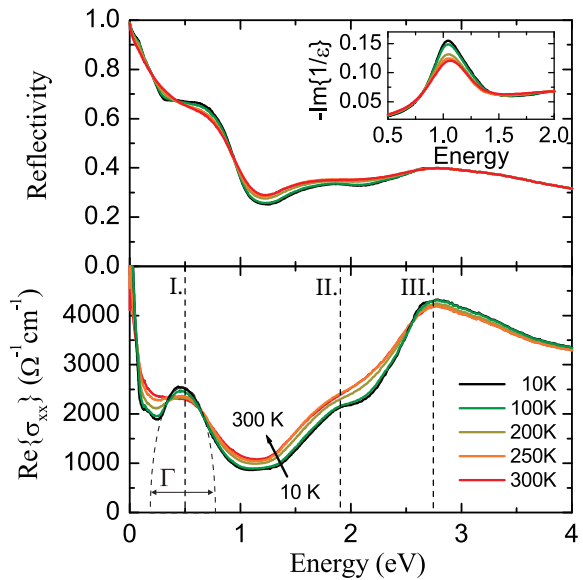}
\caption{(Color online) Upper and lower panel: The reflectivity and the optical
conductivity spectra of CuCr$_2$Se$_4$ at various temperatures.
Three main features are indicated by dashed lines. The first charge
transfer excitation (Se 4p$\rightarrow$Cr 3d) appears at
$E$$=$$2.75$\,eV with a low-energy shoulder at $E$$=$$1.9$\,eV due to on-site d-d transition of the chromium ions. In
addition to the low-energy metallic peak, a strong transition is
present in the mid-infrared region (at $E$$=$$0.5$\,eV) with a
characteristic width of $\Gamma$$\approx$$0.5$\,eV. In the inset the
maximum of the loss-function indicates the plasma frequency at
$\hbar\omega_{pl}$$\approx$$1$\,eV.} \label{fig1}
\end{figure}

\section{RESULTS AND DISCUSSION}

The temperature dependence of the reflectivity and the diagonal
optical conductivity spectra is shown in Fig.~\ref{fig1}. We
identified the different contributions to the optical conductivity
as follows. The first charge transfer peak is centered around
E=2.75\,eV, that we assign to Se 4p $\rightarrow$ Cr 3d transition
in agreement with former optical data on a broad variety of
chromium spinel oxides and chalcogenides. \cite{Ohgushi08} It has
a low-energy shoulder located at E=1.9\,eV which likely originates
from the on-site chromium d-d transition since this structure is
common for chromium spinels insensitive to the change of the other
cation. \cite{Ohgushi08} This originally dipole-forbidden
transition becomes allowed by the hybridization with the ligand,
which results in the fairly small oscillatory strength. The
spectral structures become distinct as the temperature decreases.
In the low-energy region ($E$$\lesssim$$0.1$\,eV) metallic
conductivity appears. Although it does not closely follow a
Drude-like behaviour -- the scattering rate is estimated to be
$\gamma\approx0.03$\,eV at T=10\,K -- the small residual
resistivity\cite{Ohgushi08} $\rho_o=10$\,$\mu\Omega$cm is not
typical of bad metals with strongly correlated d band. Between the
low-energy metallic term and the first charge transfer excitation,
a distinct peak appears at $E$$=$$0.5$\,eV. At low temperatures it
becomes clearly distinguishable from the excitation of the free
carriers. In spite of its closeness to the metallic continuum it
becomes sharp at low temperatures characterized by a width of
$\Gamma$$\approx$$0.5$\,eV width. In the inset of Fig.~\ref{fig1}
the maximum of the loss-function signals a plasma frequency of
$\hbar\omega_{pl}\approx$1\,eV, which only increase by 4$\%$ as
the temperature decrease to T=10\,K. This implies merely tiny
changes in the carrier concentration as a function of temperature.

\begin{figure}[t!]
\includegraphics[width=2.8in]{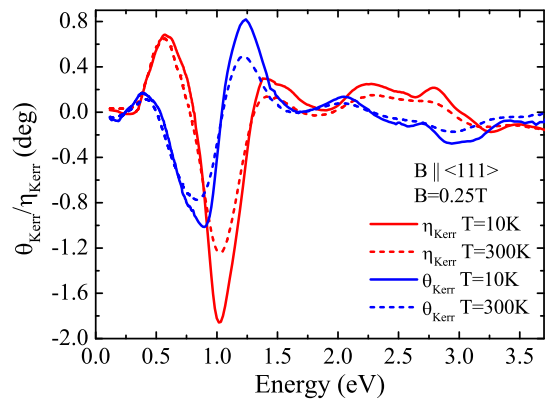}
\caption{(Color online) The spectra of the magneto-optical Kerr
parameters at room temperature and T=10\,K. The Kerr
ellipticity shows a peak at $E$$=$$1$\,eV due to the plasma resonance
and its maximal value increase to $\eta_{Kerr}$$=$$1.9^{o}$ as the
temperature decrease to T=10\,K corresponding to the sharpening of
the plasma edge. The Kerr rotation reaches the value $\theta_{Kerr}$$=$$1.2^{o}$ in the same region.} \label{fig2}
\end{figure}

The magneto-optical Kerr spectra are presented in Fig.~\ref{fig2}
at room temperature and at the lowest temperature $T$$=$$10$\,K.
These results are in good agreements with Kerr spectra previously
reported at room-temperature for $E$$>$$0.6$\,eV. \cite{Reim91}
The MOKE signal reaches its maximum around $E$$=$$1$\,eV, where
Kerr ellipticity exhibits a peak while Kerr rotation has a
dispersive line shape with the maximal values of
$\eta_{Kerr}$=1.9$^o$ and $\theta_{Kerr}$=-1$^o$, respectively.
Although the magnetization is almost constant below room
temperature, the MOKE is enhanced by 45$\%$ down to $T$$=$$10$\,K.

\begin{figure}[t!]
\includegraphics[width=3in]{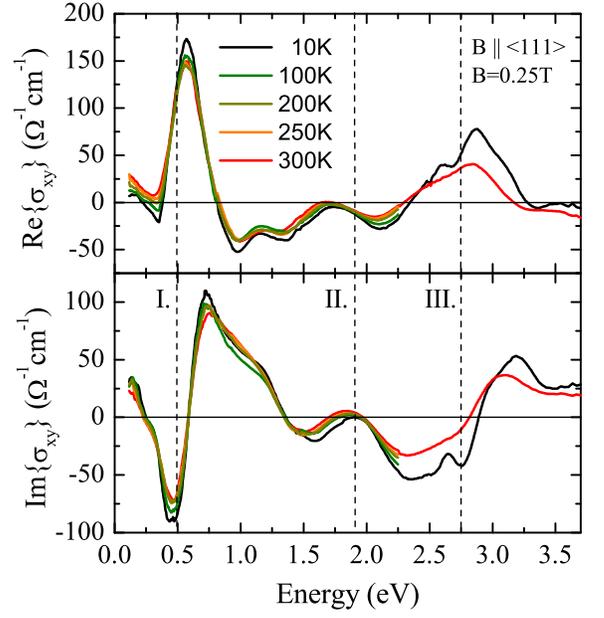}
\caption{(Color online) The off-diagonal conductivity determined
from the magneto-optical Kerr spectra. The plasma edge resonance
observed in the Kerr effect is cancelled out, which indicate that
the Kramers-Kronig transformation was performed properly. The
dashed lines with labels indicates the same transition like in
Fig.~\ref{fig1}. The transition at $E$$=$$0.5$\,eV shows large while the Se 4p$\rightarrow$Cr 3d charge transfer excitation has also considerable magneto-optical activity.} \label{fig3}
\end{figure}

From the complex Kerr angle, we have calculated the off-diagonal conductivity according to the
following relation:
\begin{equation}
\Phi_{Kerr}=\theta_{Kerr}+i\eta_{Kerr}=-\frac{\sigma_{xy}}{\sigma_{xx}\sqrt{1+\frac{4\pi i}{\omega}\sigma_{xx}}},
\label{sxy_formula}
\end{equation}
where $\sigma_{xx}$ and $\sigma_{xy}$ are the elements of the
complex optical conductivity tensor. The Kerr rotation and
ellipticity describe the phase-shift and the intensity difference,
respectively, between left and right circularly polarized light
upon normal-incidence reflection from a magnetic surface. The
corresponding results are presented in Fig.~\ref{fig3}.

In spite of the large MOKE around $E$$=$$1$\,eV, neither the
off-diagonal nor the diagonal conductivity show any specific
optical excitation in this energy region. The large enhancement of
the MOKE signal corresponds to the plasma resonance at
$\hbar\omega_{pl}\approx$$1$\,eV; it is caused by the strong
minimum of the denominator of Eq.~\ref{sxy_formula} rather than by
an increase of the off-diagonal conductivity.\cite{Reim91,Feil87}
The almost perfect cancelation of this resonance in the
off-diagonal conductivity indicates the properness of the
Kramers-Kronig transformation for the reflectivity. When the
optical excitations are broad compared to the magnetically induced
splitting of these transitions for the two circular polarizations,
the Kerr parameters $\theta_{Kerr}$ and $\eta_{Kerr}$ are
proportional to the derivative of the reflectivity:
\begin{eqnarray}
\eta_{Kerr}=\frac{1}{2}\frac{r_+^2-r_-^2}{r_+^2+r_-^2} \propto \frac{1}{R(E)}\frac{\partial R(E)}{\partial E} \nonumber\\
\theta_{Kerr}=\frac{1}{2}(\phi_+-\phi_-) \propto \frac{\partial
\phi(E)}{\partial E},
\end{eqnarray}
where $\tilde{r}_\pm$=r$_\pm$e$^{\phi_\pm}$ are the Fresnel
coefficients for the right and left circularly polarized photons
and $R(E)$$=$$(r_+^2+r_-^2)/2$ and $\phi (E)$ are the reflectivity
and the corresponding phase. The sudden decrease of the
reflectivity near the plasma edge, which generate the large MOKE
signal, is sensitive to the slope of the reflectivity which
becomes steeper as the life-time increases toward low
temperatures, causing considerable temperature dependence in the
region of the plasma resonance.

\begin{figure}[t!]
\includegraphics[width=3in]{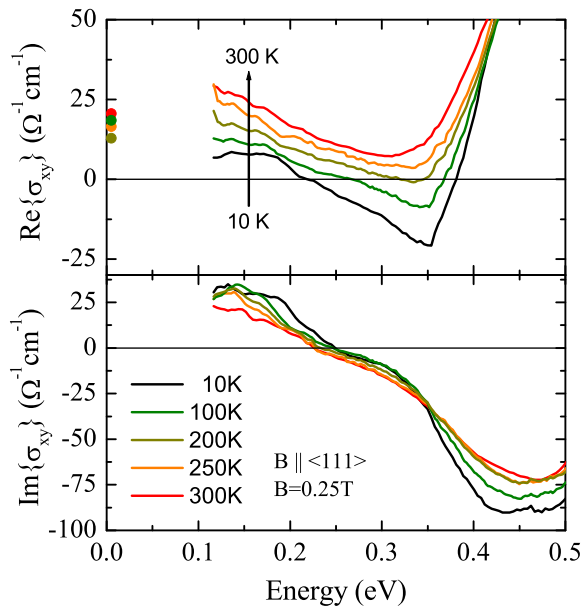}
\caption{(Color online) The low-energy part of the off-diagonal
conductivity. The dc values are reproduced from Hall resistivity
measurement performed in the same magnetic field by Lee et al.\cite{Ong04}} \label{fig4}
\end{figure}

The real part of the off-diagonal conductivity (shown in
Fig.~\ref{fig3}) is dominated by two main structures, namely a
broader hump around $E$$=$$2.75$\,eV and a resonance-like peak centered
at $E$$=$$0.5$\,eV. The dispersive line-shape in the imaginary part of
the off-diagonal conductivity and also the corresponding large values
-- Im$\{\sigma_{xy}\}$=53\,$\Omega^{-1}cm^{-1}$ and
Im$\{\sigma_{xy}\}$=110\,$\Omega^{-1}cm^{-1}$ at $T$$=$$10$\,K,
respectively -- are suggestive of parity allowed transitions,
which is reasonable for the Se 4p$\rightarrow$Cr 3d charge
transfer transition at $E$$=$$2.75$\,eV. The magnitude of the low-energy
part of the off-diagonal conductivity enlarged in Fig.~\ref{fig4}.
is very close to that of the dc Hall effect obtained in the same
magnetic field,\cite{Ong04} except for $T$$=$$10$\,K where
$\sigma_{Hall}$$=$$300$\,$\Omega^{-1}cm^{-1}$. As the temperature
decreases the low-energy tail of the real part is considerably
reduced in contrast to the temperature independent behavior of the
magnetization. This may indicate that a non-perturbative treatment
of the spin-orbit coupling is also necessary to describe the low-energy off-diagonal
conductivity as it was formerly proposed for the dc anomalous Hall effect.\cite{Ong04,Fang07}

The relations between the elements of the optical conductivity tensor and the underlying
microscopic optical processes are described by the Kubo formula:
\begin{eqnarray}
Re\{\sigma_{xx}\}=\frac{\pi e^2}{2m^2V\hbar\omega}\sum_{i,f}
[1-f(\varepsilon_f)]f(\varepsilon_i) \{|\langle
f|\Pi_+|i\rangle|^2+ \nonumber\\
+ |\langle f|\Pi_-|i\rangle|^2\}
[\delta(\omega_{fi}-\omega)+\delta(\omega_{fi}+\omega)],
\nonumber\\
Im\{\sigma_{xy}\}=\frac{\pi e^2}{4m^2V\hbar\omega}\sum_{i,f}
[1-f(\varepsilon_f)]f(\varepsilon_i) \{|\langle
f|\Pi_+|i\rangle|^2- \nonumber\\
- |\langle f|\Pi_-|i\rangle|^2\}
[\delta(\omega_{fi}-\omega)+\delta(\omega_{fi}+\omega)],\label{Kubo_formula}
\end{eqnarray}
where $\Pi_\pm=\Pi_x\pm i\Pi_y$ are the momentum operators in
circular basis. The real part of the diagonal optical conductivity
is proportional to the joint density of states for the occupied
and unoccupied states multiplied by the electric dipole matrix
elements, therefore, it describes the absorption of light for
left and right circularly polarized photons in average. On the other hand, the off-diagonal conductivity is the
difference between the absorption spectra corresponding the two
circular polarizations. Contributions from electric dipole processes to off-diagonal optical conductivity are remarkable in ferromagnetic materials -- due to the orbital magnetization induced by the
spontaneous spin polarization via the spin-orbit interaction -- similarly to the anomalous Hall effect in the dc limit.

\begin{figure}[t!]
\includegraphics[width=3.3in]{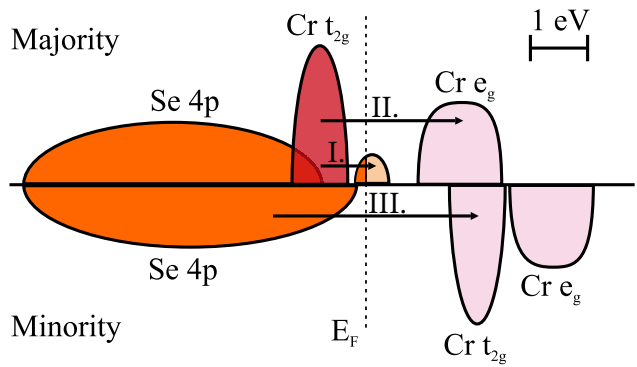}
\caption{(Color online) The schematic structure of the density of
states as it determined by density functional calculations.
\cite{Antonov99,Sarma07,Butler08} The chromium d-band is split by
the cubic crystal field to a t$_{2g}$ and an e$_g$ band. The hybridization between the selenium and chromium induce a
gap just below the Fermi energy and a holes appears in the
majority spin channel. The arrows indicate the excitations observed both
in the diagonal and off-diagonal conductivity spectra with the same labels as used in Fig.~1 and 3.} \label{fig5}
\end{figure}

To understand the origin of the optical excitations, the results
of the recent band structure calculations\cite{Antonov99, Sarma07,
Butler08} are summarized schematically in Fig.~\ref{fig5}. The
chromium d-band is split by the cubic crystal field to a t$_{2g}$
and an e$_g$ band. The selenium 4p and the copper 3d states are
fully mixed with each other. Furthermore, the hybridization
between the selenium 4p and the chromium t$_{2g}$ induces a gap
just below the Fermi energy in the majority spin channel. As a
consequence, states from the Se 4p band are shifted above the
Fermi level (referred to as "hump in the density of states" in the
introduction), thus holes appears in the majority spin channel.
The main optical transitions observed in the experiments are also
indicated in the figure.

Our optical and magneto-optical study confirms the results of the
band structure calculations both in the close vicinity of the
Fermi energy and on a few eV large scale. In agreement with the
theoretical results, we explain the transition at $E$$=$$0.5$\,eV
as excitations trough the hybridization induced gap. The high
oscillator strength is due to the parity difference between the
initial and the final states. The large off-diagonal conductivity
is possibly owing to the strong spin-orbit coupling for the
delocalized electrons with strong selenium character
(E$_{SO}$$\approx$$0.5$\,eV) and the highly spin polarized states
in the $\sim$1\,eV vicinity of the Fermi level.\cite{Antonov99,
Butler08} The first charge transfer excitations around
$E$$=$$2.75$\,eV, with a remarkable oscillator strength and
magneto-optical activity are attributed to Se 4p$\rightarrow$Cr 3d
transition, while the excitation at $E$$=$$1.9$\,eV is assigned to
on-site d-d transitions of chromium ions. These are in overall
agreement with the numerical calculations, although the transition
energies are somewhat higher in the experiment. \cite{Butler08,
Sarma07, Antonov99} The contribution of the d-d transition to the
off-diagonal conductivity is small compared to that of the charge
transfer transition. Besides the reduced oscillator strength of
the d-d transition due to its dipole forbidden nature, it is
likely caused by the fairly small spin-orbit coupling for chromium
(E$_{SO}$$=$$0.09$\,eV). \cite{Antonov99}

The measurement of the off-diagonal conductivity greatly helps to
extract the optical transition found around $E$$=$$0.5$\,eV since
its large magneto-optical activity dominates over the contribution
from the metallic charge carriers (damped cyclotron resonance)
contrary to the case of the diagonal conductivity.

\section{CONCLUSIONS}

We have measured the reflectivity and magneto-optical Kerr effect
over a broad energy range (E=0.08-26\,eV and E=0.1-4\,eV,
respectively) at various temperatures down to T=10\,K and
evaluated the elements of the optical conductivity tensor.
The diagonal and off-diagonal optical conductivity spectra determined from the experiments are
consistent with the results of former band structure
calculations\cite{Antonov99,Sarma07,Butler08} over the whole
energy region. At low energies a metallic peak is present, while the
E$\gtrsim$2\,eV region is dominated by the first charge transfer
transition Se 4p$\rightarrow$Cr 3d. Moreover, we have observed
a distinct optical transition around $E$$=$$0.5$\,eV that we attribute to excitations through the Se-Cr hybridization induced gap. This transition has huge magneto-optical activity due
to the high spin polarization in the $\sim$1\,eV vicinity of the
Fermi level and its parity-allowed nature. The corresponding sharp feature in the off-diagonal conductivity dominates over the contribution from the metallic electrons. Large enhancement of the Kerr effect was also observed around the plasma edge.

This work was supported by a Grant-In-Aid for Scientific Research,
MEXT of Japan, and the Hungarian Research Funds OTKA PD75615,
NK72916, Bolyai 00256/08/11.


\begin{references}
%
\bibitem{Landolt}Landolt-B\"{o}rnstein, \textit{Magnetic and other properties of oxides and related compounds}, Vol III. part 12b, Springer, Berlin (1988) and the
references there.
%
\bibitem{Reim91}H. Br\"{a}ndle, J. Schoenes, P. Wachter, F. Hulliger and W. Reim, J. Magn. Magn. Mat. \textbf{93}, 207 (1991).
%
\bibitem{Ramirez97}A. P. Ramirez, R. J. Cava and J. Krajewski, Nature \textbf{386}, 156 (1997).
%
\bibitem{Loidl05}J. Hemberger, P. Lunkenheimer, R. Fichtl, H.-A. Krug von Nidda, V. Tsurkan and A. Loidl, Nature \textbf{434}, 364 (2005).
%
\bibitem{Antonov99}V. N. Antonov, V. P. Antropov, B. N. Harmon, A. N. Yaresko and A. Ya. Perlov, Phys. Rev. B \textbf{59}, 14552 (1999).
%
\bibitem{Sarma07}T. Saha-Dasgupta, Molly De Raychaudhury and D. D. Sarma, Phys. Rev. B \textbf{76}, 054441 (2007).
%
\bibitem{Butler08}Y.-H. A. Wang, A. Gupta, M. Chshiev and W. H. Butler, Appl. Phys. Lett. \textbf{92}, 062507 (2008).
%
\bibitem{Lotgering67}F. K. Lotgering and R.P. Stapele, Solid State Commun. \textbf{5}, 143 (1967).
%
\bibitem{Colominas66}C. Colominas, Phys. Rev. \textbf{153}, 558 (1966).
%
\bibitem{Goodenough69}J. B. Goodenough, J. Phys. Chem. Solids \textbf{30}, 261 (1969).
%
\bibitem{Kimura01}A. Kimura, J. Matsuno, J. Okabayashi, A. Fujimori, T. Shishidou, E. Kulatov and T. Kanomata, Phys. Rev. B \textbf{63}, 224420 (2001).
%
\bibitem{Ohgushi08}K. Ohgushi, Y. Okimoto, T. Ogasawara, S. Miyasaka and Y. Tokura, J. Phys. Soc. Jpn.
\textbf{77}, 034713 (2008).
%
\bibitem{Sato81}K. Sato, Jpn. J. Appl. Phys. \textbf{20}, 2403 (1981).
%
\bibitem{Feil87}H. Feil and C. Haas, Phys. Rev. Lett. \textbf{58}, 65 (1987).
%
\bibitem{Ong04}Wei-Li Lee, S. Watauchi, V. L. Miller, R. J. Cava and N. P. Ong, Science \textbf{303}, 1647 (2004).
\bibitem{Fang07}Y. Yao, Y. Liang, D. Xiao, Q. Niu, Shun-Qing Shen, X. Dai, and Z. Fang, \prb \textbf{75}, 020401(R) (2007).
\end{references}
\end{document}